# SIMPLIFYING URBAN DATA FUSION WITH BigSUR

Authors:

**Tom Kelly & Niloy J. Mitra**

Institution:

UNIVERSITY COLLEGE LONDON

## ABSTRACT

Our ability to understand data has always lagged behind our ability to collect it. This is particularly true in urban environments, where mass data capture is particularly valuable, but the objects captured are more varied, denser, and complex. Captured data has several problems; it is unstructured (we do not know objects are represented by the data), contains noise (the scanning process is often inaccurate) and omissions (it is often impossible to scan all of an area). To understand the structure and content of the environment, we must process the unstructured data to a structured form.

BigSUR[1] is an urban reconstruction algorithm which fuses GIS data, photogrammetric meshes, and street level photography, to create clean representative, semantically labelled, geometry. However, we have identified three problems with the system i) the street level photography is often difficult to acquire; ii) novel façade styles often frustrate the detection of windows and doors; iii) the computational requirements of the system are large, processing a large city block can take up to 15 hours.
In this paper we describe the process of simplifying and validating the BigSUR semantic reconstruction system. In particular, the requirement for street level images is removed, and greedy post-process profile assignment is introduced to accelerate the system. We accomplish this by modifying the binary integer programming (BIP) optimization, and re-evaluating the effects of various parameters.

The new variant of the system is evaluated over a variety of urban areas. We objectively measure mean squared error (MSE) terms over the unstructured geometry, showing that BigSUR is able to accurately recover omissions from the input meshes. Further, we evaluate the ability of the system to label the walls and roofs of input meshes, concluding that our new BigSUR variant achieves highly accurate semantic labelling with shorter computational time and less input data.

## INTRODUCTION

### Procedural Extrusions

We aim to fit a semantically meaningful parameterized model to real world mesh and cartographic data. In the process, we remove the noise from the mesh, creating a simplified "clean" model. To achieve this we decompose a building to a horizontal 2D footprint, and a profile associated with every edge. BigSUR used such a parameterization - *procedural extrusions*[13] *(PEs)*. This decomposition results in a model which can label the 2D extent of the building over the floorplan, and identify the walls and roofs from the profiles.

PEs are mathematically elegant roofs – they are built using a variant of the straight skeleton[2], a geometric construct which has the property that wherever a raindrop lands on the roof, if it continues downhill, it will always reach a gutter at the edge of the roof. The issue when modeling with PEs is to determine where the gutters should be. To do this is a reliable way, such that the accompanying profiles can accurately represent the roof geometry, is a challenge – the top of the roof crest may be a long distance from the gutter above the building boundary.

The BigSUR method performs such a decomposition to PEs, with the aid of street-level photography to determine locations for joins between buildings – likely locations for gutters. However, such photographs are expensive to collect or label, and it may not be possible to photograph all facades of a building (i.e. interior courtyards). Here we wish to simplify the BigSUR method, and illustrate how it can work without such images.

**Related Work**

Urban reconstruction is a large subject area spread between subjects as diverse as geomatics, computer graphics, and geometry. For a more complete survey of the work, we refer the read to the reports by Wang et al.[3] and Musialski et al.[4] However, here we will satisfy ourselves with a short tour between two extremes of reconstruction – *mesh modeling* and *primitive modeling*.

*Mesh modeling* involves creating vertices and faces of an arbitrary mesh in such a way to model the target data. These approaches are very flexible, and can construct arbitrary data. However, they often lack prior knowledge of the urban environment, and so their worst-case results do not resemble architectural structures. A classic example is Screen Poisson Surface Reconstruction[5], which can operate on a wide variety of data but often results in "blobby" artefacts which are better suited to organic objects than the built environment. Another technique is *dual contouring*, a technique which has been modified for urban reconstruction by using the method in 2.5 dimensions to create only building roofs[6]. However, because the roofs can take arbitrary shapes the worst-case quality can be poor. Salinas et al.[7] use detected faces and edges to regularise a mesh. The work is able to elegantly simplify manifold meshes. However, because they do attempt to understand the semantics of the geometry, they are unable to apply urban priors such as ensuring walls are vertical.

*Primitive modeling* involves arranging pre-existing objects to reconstruct the urban environment. Working at this higher level results in systems that can perform can create very convincing urban outputs from very noisy data. However, they are only able to model objects that exist in their libraries. A classic example is presented by Vanegal et al. [8]; they reconstruct environments using only cuboids, this guarantees that the results are well formed and manifold, but cannot represent sloped roofs. Verma et al.[9] use a larger set of parameterised primitives which contains structures with sloped and flat roofs. Finally, Edelsbrunner introduce a solid roof primitive[10] that is able to create impressively complex roofs. However, the choice of primitive limits the results, for example flat roofs, or buildings with arbitrary footprints cannot be represented.

Procedural extrusions lie between these extremes of mesh and primitive modeling, offering many of the advantages of both approaches. As we hope to show in the following PEs, like mesh modeling systems, can be quickly parameterised to real world data, and are not limited to a set of known primitives. In addition, they have many of the advantages of primitive modeling – PEs are easy to edit, are guaranteed to create watertight geometry, and contain semantically valuable information. Finally, PEs have an excellent worst case reconstruction, as given a strong urban profile prior, their result is usually resembles *architecture*.

**METHOD**

To the above ends, we simplify the method of BigSUR by modifying the central optimization problem. Broadly, we achieve this by removing the complications that come with photographic data (the *building-facades*) and profile assignment. This remove terms $O_3$, $O_4$, $O_5$, and $O_6$ from the previously published optimization, and adds an additional post-processing step to assign profiles. We continue to describe how we find the inputs to the optimization, the way in which the optimization is formulated, the profile postprocessing, and finally how we create the mass model.

## Computing the optimization inputs

The inputs to our system are 2D GIS (cartographic) building footprints, and a noisy 3D photogrammetric mesh. Here we describe how these are processed to create the input to the optimization - a set of *sweep-edges*. Sweep-edges are 2D lines in the ground-plane giving the approximate locations of the vertical walls.

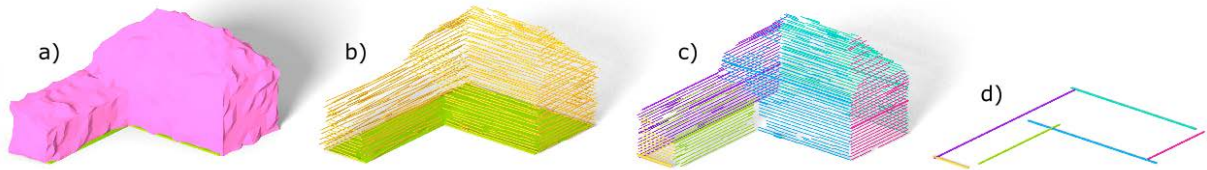

*Figure 1. Computing the sweep-edges.*

To compute the sweep-edges, we take the input GIS footprints and 3D mesh (Figure 1, a). At vertical intervals of 0.2m we slice the mesh horizontally to create a large number of horizontal lines (b). These lines are aligned to nearby edges in the GIS footprint. We continue to cluster these lines into different prominent directions (c), each representing a side of the building with a single profile. We discard any clusters with an associated mesh area below a certain limit, $\gamma$, typically $10m^2$. The base of these profiles is projected onto the ground plane to create the final sweep-edges. These sweep-edges should follow the gutters of the roof over the structure.

The sweep-edges are the result of heavy processing over noisy data. Because of this, they do not form closed footprints. Some may be missing, others may interpenetrate each other, or they may not intersect where we expect them to (i.e. the corners of buildings). For these reasons we perform an optimization to reconstruct plausible footprints from the sweep-edges.

## Optimization Terms

Given the set of sweep-edges we describe here how we find a set of watertight building footprints for the block from the sweep-edges. We first fracture the ground-plane into a large number of polygons, then formulate an optimization which assigns a footprint-label each polygon.

The sweep-edges are used to fracture the ground-plane. Starting with the longest, each sweep-edge is inserted into the plane, fracturing the plane into polygons (Figure 2 a-d). To remove further complexity from the following optimization, we can perform an inside-outside segmentation using the GIS footprint and mesh-height to discard those areas outside of the building (Figure 2 e-f).

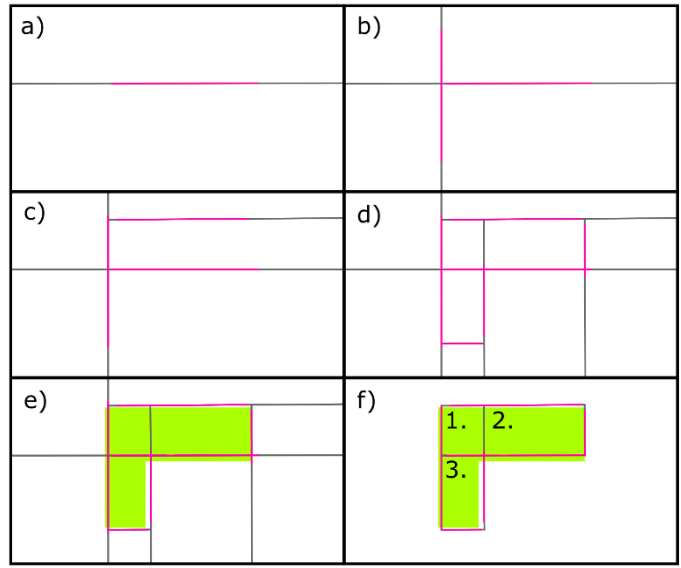

*Figure 2. Fracturing the ground plane. a-d) Sweep-edges (pink) and their continuations (grey) are used to divide the plane into many polygons. The GIS data (e, green) is used to identify polygons inside the building. In this simple example the fracturing results in polygons 1-3 (f).*

We now wish to label polygons which belong to the same footprint. Typically, there will be many more polygons than desirable footprints, and we must find which polygons we might combine to a single footprint. Following BigSUR, we define error terms $O_1$, $O_2$, and $O_7$:

$$O_1 = \sum_{e_k} \alpha \, \|e_k\| \left( \neg s_k \wedge isSweepEdge(e_k) \right) + \beta \, \|e_k\| \left( s_k \wedge \neg isSweepEdge(e_k) \right)$$

$$O_2 = \sum_{e_k} \|e_k\| \, heightDiff(e_k) \neg s_k$$

$$O_7 = \sum_{e_i, e_j} \varphi \cdot s_j \wedge s_k$$

Where:

| | |
|---|---|
| $e_k$ | Each edge in each polygon |
| $\|e_k\|$ | The length of edge $e_k$ |
| $\alpha, \beta$ | Weights that control the balance between over and under segmentation |
| $s_k$ | Binary value, 1 if $e_k$ is part of an output footprint, otherwise 0 |
| $isSweepEdge(e_k)$ | Binary value, 1 if $e_k$ is a sweep edge or continuation edge (see Figure 2) |
| $heightDiff(e_k)$ | The height difference of the noisy mesh across $e_k$ |
| $\varphi$ | A large penalty term: $0.5 \sum_{e_k} \|e_k\|$ |
| $e_i, e_j$ | Pairs of polygon edges that are closer than 2m, or are adjacent and form and angle of less than 30 degrees. Such geometry is undesirable. |

We search for a labelling such that $O_1 + O_2 + O_7$ is minimized. In the above, this means searching the valid values of $s_k$ for the most desirable solution.

We represent the problem as a binary integer programming (BIP) task and solve using the Gurobi[11] library to assign a footprint-index to each face. Given the resulting indices we can join adjacent polygons with the same index to create our set of footprints. The resulting footprints can be quite complex, and may contain holes.

**Assigning Profiles as a Postprocess**

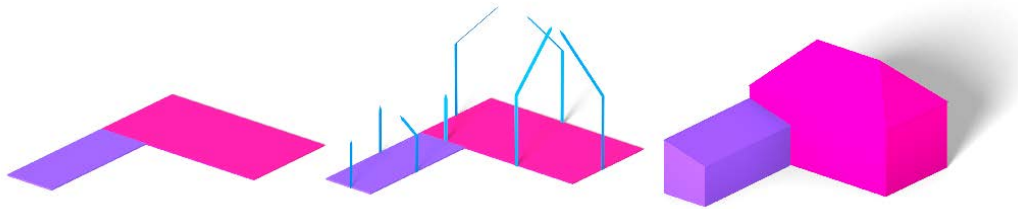

*Figure 3. Left: the output of the optimization is a set of building footprints. Center: profiles are assigned as part of our post processing. Right: Using procedural extrusions we combine the profiles and the footprints.*

Given the output footprints from the optimization (Figure 3, left), here we describe how we assign profiles to each edge (Figure 3, centre).

As illustrated in Figure 4, we take every edge of the footprints, and traverse the input mesh to find a set of noisy profiles (a). By starting at sampled points along the edge, we slice the mesh perpendicular to the edge, and climb the slice until we cannot find a higher point (b). We can then use standard techniques to clean and merge these polylines into a single clean profile (c). At this point we apply our strong strong urban prior to the shape of the profiles – we expect a vertical wall bellow one or more sloped roof pitches.

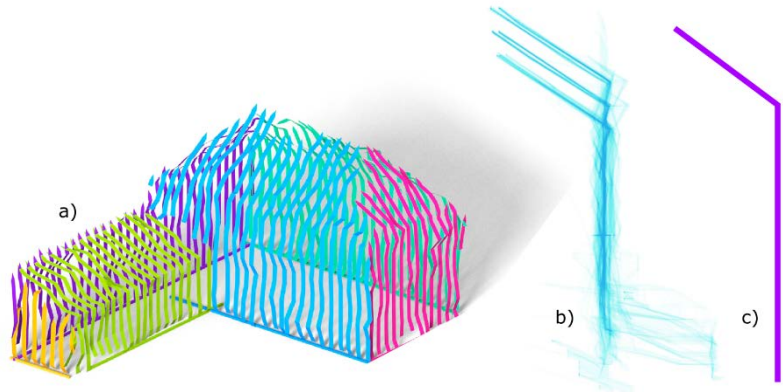

*Figure 4. a) Given footprints, we can slice the mesh to find many profiles.
b) one footprint's profiles. c) the cleaned profile.*

**Creating the Mass Model**

Give the footprints from the optimization, and the found profiles (Figure 3, centre) we can continue to compute the final building masses. We use the *campskeleton*[12] implementation of the *procedural extrusion*[13] system to achieve this. The right-hand panel of Figure 3 shows an example of such a mass model. PEs vertically extrude a building's 2D footprint along the given profiles. PEs can be subject to geometry run-away; for example, when only vertical profiles are found the footprint is extruded to infinity. To avoid this, we use a maximum extrusion limit computed from the input mesh.

**Implementation**

We have released an implementation of this method, *chordatlas*[14]. In particular, the modifications in this paper can be activated by using the "use greedy profiles" option in the settings menu. The system allows the above optimization to be run, results may be exported, and the resulting footprint and profile representation may be edited. The source code contains details on the many hundreds of practical details required to replicate this work.

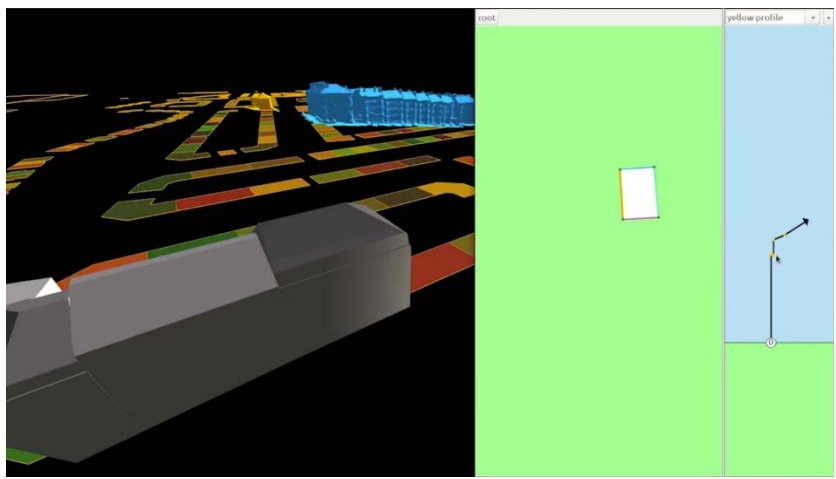

*Figure 5. Our* chordatlas *urban data fusion platform. The 3D view on the left shows a photogrammetric mesh (blue), GIS footprints (orange), and the output model (grey). On the right
we see an editor for the plan and profile of one footprint.*

## PARAMETER EXPLORATION

In the above we introduced several parameters, such as $\alpha, \beta, \gamma$. Here we explore the effect of these parameters on our output models.

### Over- or Under-Segmentation

The constants $\alpha$ and $\beta$ control the under- and over-segmentation in the system. $\alpha$ controls the penalty associated with a sweep edge that is not part of a footprint, while $\beta$ controls the penalty given to lengths of a footprint which are not associated with a sweep edge.

We explored 4 different combinations of these values, and the results are shown in Figure 6. We observe that a high $\beta$ leads to an under-segmentation (top), while a high $\alpha$ leads to over segmentation (bottom). We present a vertical error plot, and average mean-squared error for several parameter combinations. Given these results we use $\alpha = 40 \; and \; \beta = 60$ for further experiments.

### Profile Quality

We found it instructive to explore the consequence of more or less detailed profiles. By tweaking our profile simplification algorithm we could quickly change the visual complexity of a block. Our results are shown in Figure 7. Simple (a single vertical line) profiles, create a bounding-volume-like representation. High complexity profiles created very realistic buildings. Moderately complex profiles significantly reduced the polygon count of our models while retaining a good visual quality. We used the high profiles for all other results presented.

### Sweep-edge Area Threshold

The final parameter we investigated was the minimum-area threshold for a sweep-edge to be used, $\gamma$. We varied this parameter between 10 and 90m$^2$. The results are shown in Figure 8; the first result, a, shows very high error because the boundaries between the terraced houses are not detected. As these edges are progressively introduced, the quality of the result increases.

We note that as the number of sweep-edges increases, the error falls, the visual realism increases and the optimisation run-time increases. For the results in this paper, we used a value of $\gamma = 10$ m$^2$ with the exception of the very large New-York block, for which we used a threshold of $\gamma = 50$ m$^2$.

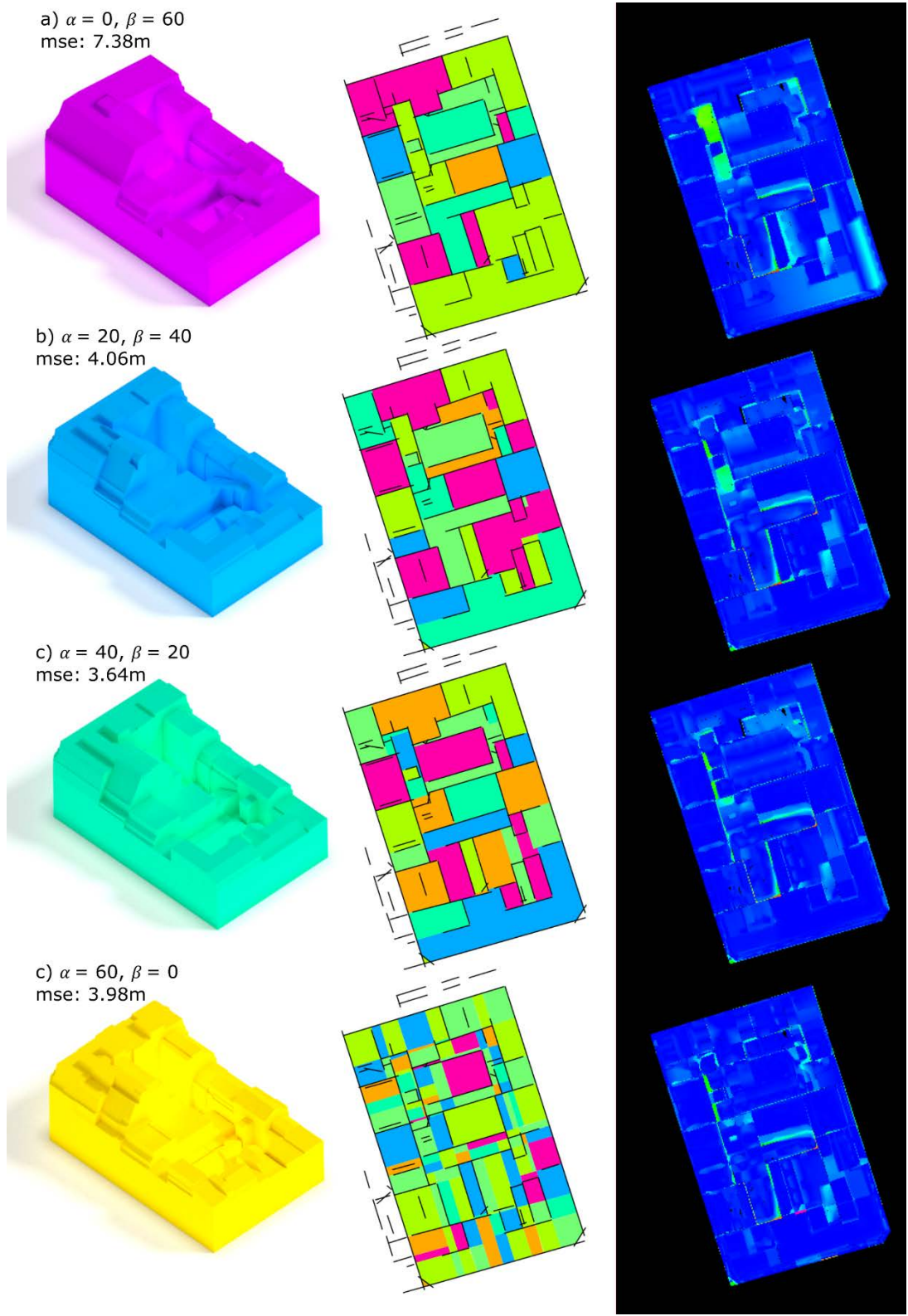

Figure 6. Rows: effects of α and β the resulting segmentation. Input mesh as Figure 7. Left: output mesh. Middle: optimized segmentation. Right: Linear error plot.

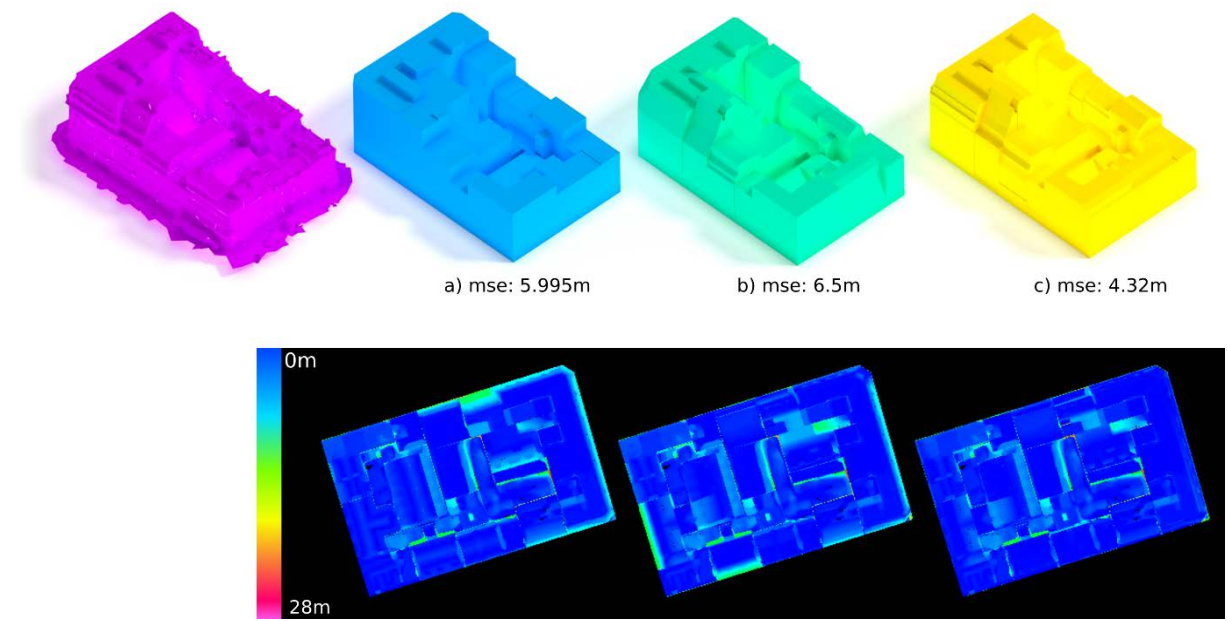

*Figure 7. Effect of profile quality on the London dataset. Purple: input Mesh. Blue: simple profiles. Green: Moderately complex profiles. Yellow: High complexity profiles. Bottom: corresponding vertical error plots.*

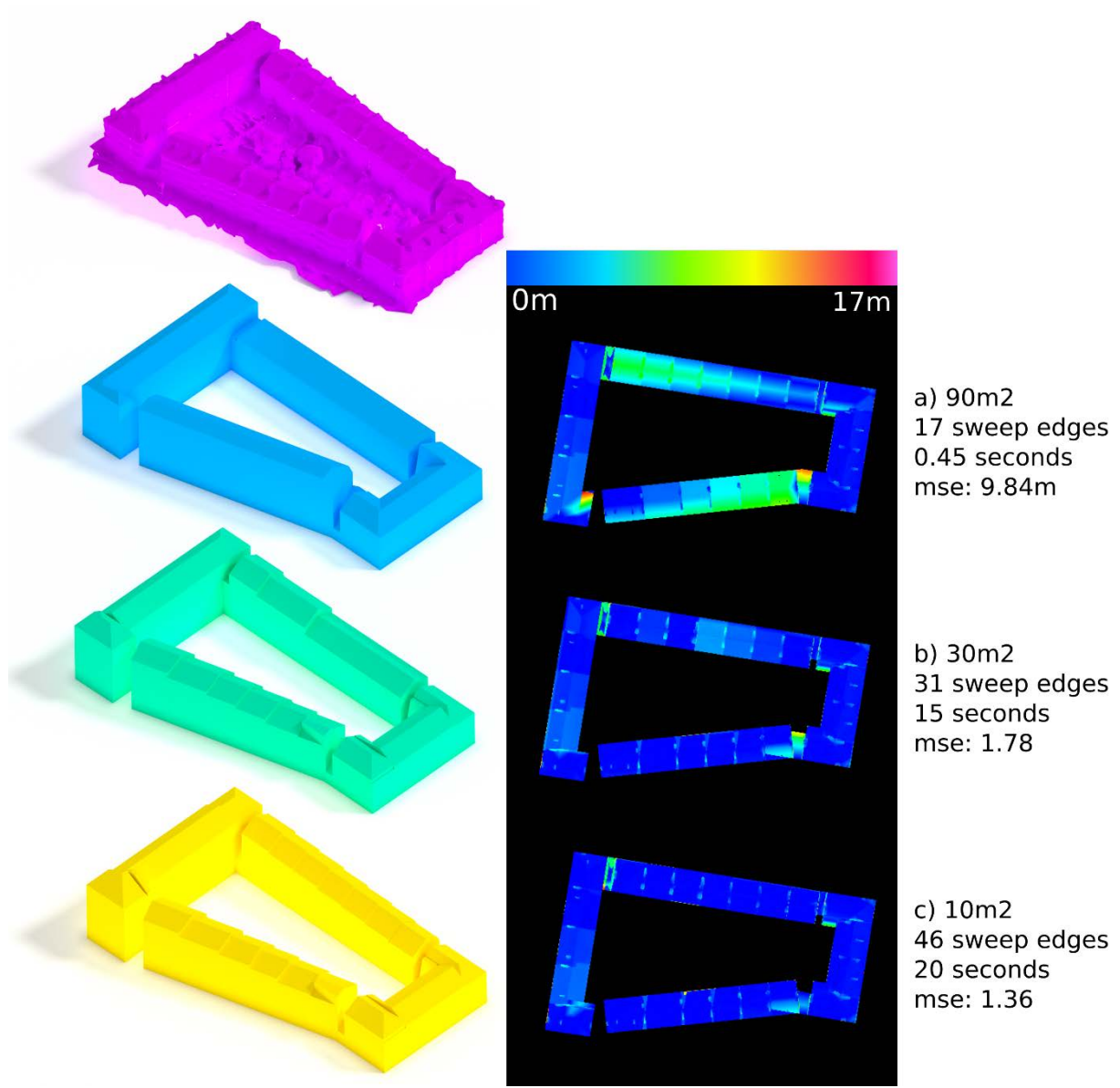

*Figure 8. The effect of the area-threshold on selecting sweep-edges. purple: the input mesh. Rows a-c) illustrate decreasing thresholds result in more accurate results (center), with lower errors (right) when compared the input meshes.*

## RESULTS

Using the parameters found in the previous section, we now present the application of simplified system to 6 real-world datasets. The inputs are a GIS footprint (either from the OS[15] for Glasgow and London results, otherwise OpenStreetMap[16]), and a photogrammetric mesh. We did not use the presented tools to edit any of the results – they are the results of our fully automated reconstruction procedure.

The results and accompanying statistics for these datasets are shown in Figure 9 and Table 1. We observe that the mean squared error varied between 0.3 and 6.0m, with the larger blocks typically having large errors due to accumulated inaccuracies as each procedural extrusion is evaluated.

Our simplifications have been successful in reducing the runtime. For example, the time to process the London dataset has been reduced from 4 hours in the BigSUR paper to 13 minutes. In other experiments on single-family homes or detached houses the optimization time was less than 0.5 seconds, allowing results at an interactive speed.

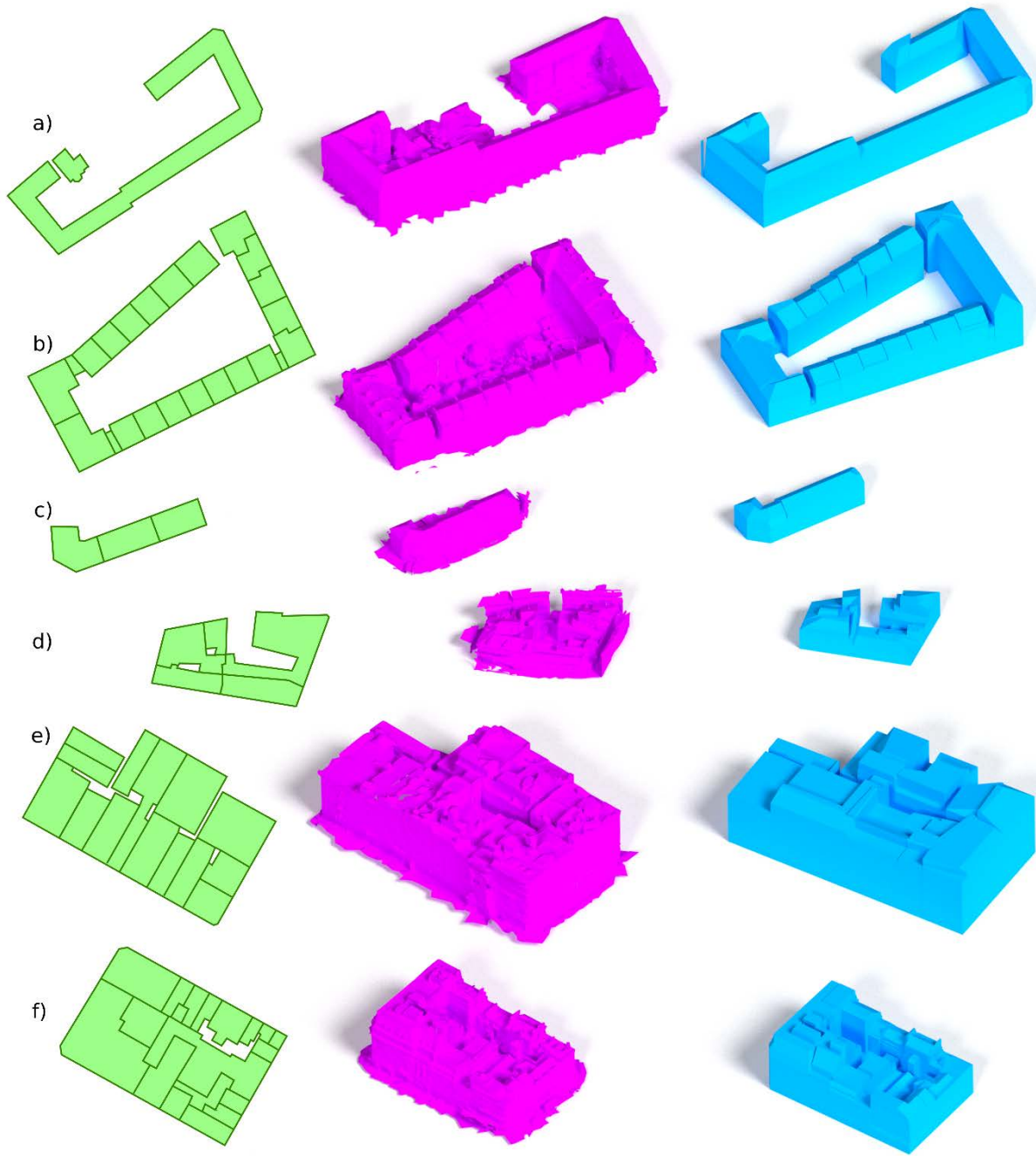

*Figure 9. Results for datasets. Green: GIS footprints. Purple: input 3D meshes. Blue: output clean meshes.*

We note that the higher, more semantic representation we construct causes errors to be expressed at this higher semantic level. For example, when a roof pitch is given an inaccurate angle, it affects the entire roof, and the roof-line of the building. Another limitation is that our polygonal representation is unable to model curves in either the footprint or the profile.

Table 1. Accompanying data for Figure 9 and 10

| Name | Label | Sweep-edges | Variables | Time (sec) | Error ($^2$m) | Lat | Long |
|---|---|---|---|---|---|---|---|
| Copenhagen | a | 37 | 2739 | 1.64 | 2.605 | 55.6616 | 12.5992 |
| Glasgow | b | 46 | 3290 | 11.2 | 1.360 | 55.8615 | -4.2011 |
| Glasgow-small | c | 10 | 466 | 0.04 | 0.348 | 55.8608 | -4.2004 |
| Madrid | d | 28 | 2832 | 6.18 | 2.075 | 40.4114 | -3.7037 |
| New York | e | 28 | 19754 | 455 | 6.05 | 40.7222 | -74.0022 |
| London | F | 70 | 12385 | 795 | 3.68 | 51.5173 | -0.1420 |

**Characteristic Profiles**

As an aside, we find it informative to characterise the different datasets by their profiles. Figure 10 shows the all the profiles for each dataset in a single plot. We can instantly observe the rectilinear profiles in New York, the mansard roofs of London's Regent street, and the similarities between the two buildings in Glasgow.

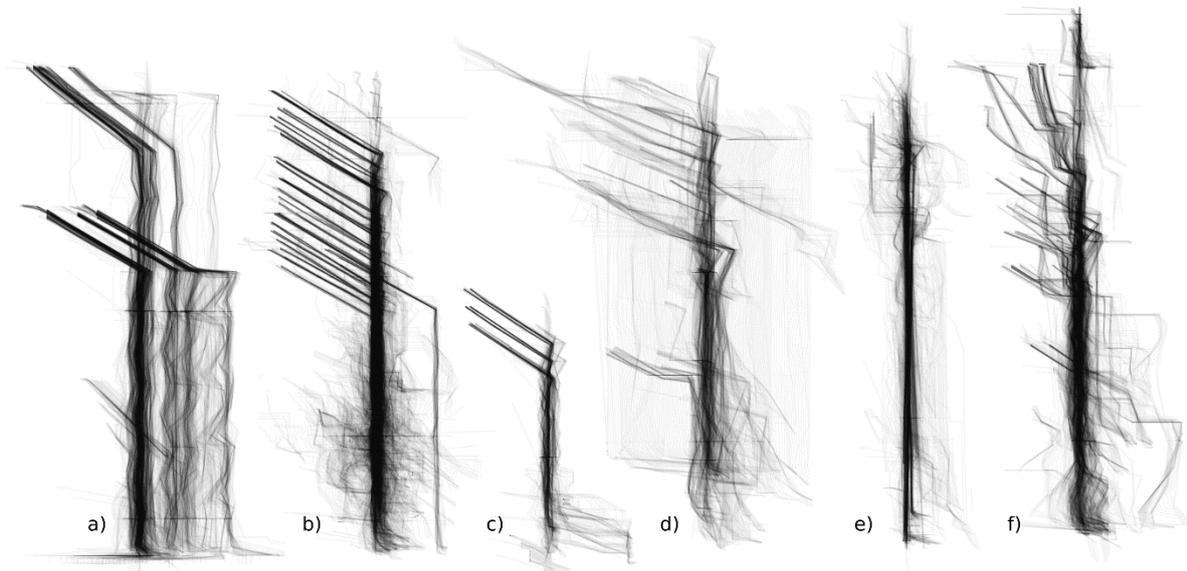

Figure 10. Profiles for Figure 9. a) Copenhagen, b) Glasgow, c) Glasgow-small, d) Madrid, e) New York, and f) London. Not to scale.

**CONCLUSIONS**

In this project we have simplified the BigSUR system, reducing the requirements for photo data, and reducing the complexity of the central optimization problem. We have searched the parameter space to gain an understanding of the possible system outputs, and used the parameters that we identified to process 6 datasets from a variety of cities.

The output models have a very high visual quality, with single walls typically presented by a single polygon, and corners of walls remaining sharp throughout processing. Further, all output models are watertight and well tessellated. Because we have decomposed the building to a footprint and a profile, we are able identify which portions of the model are wall (those with vertical profiles), and which are roofs (those with sloped profiles).